\documentclass[prd,epfs,epsfig,aps]{revtex4}
\usepackage{graphicx}%
\usepackage{dcolumn}
\usepackage{amsmath}
\begin{document}

\title{Gravitational Wave Emission from a Bounded Source:
the Nonlinear Regime}

\author{H. P. de Oliveira\footnote{Regular Associated of The Abdus
Salam International Centre for Theoretical Physics - ICTP, Trieste, Italy}}
\email{oliveira@dft.if.uerj.br}
\affiliation{{\it Universidade do Estado do Rio de Janeiro }\\
{\it Instituto de F\'{\i}sica - Departamento de F\'{\i}sica Te\'orica}\\
{\it Cep 20550-013. Rio de Janeiro, RJ, Brazil}}

\author{I. Dami\~ao Soares}
\email{ivano@cbpf.br}
\affiliation{{\it Centro Brasileiro de Pesquisas F\'{\i}sicas}\\
{\it R. Dr. Xavier Sigaud, 150. CEP 22290-180}\\
{\it Rio de Janeiro, RJ, Brazil}}

\date{\today}

\begin{abstract}
We study the dynamics of a bounded gravitational collapsing configuration
emitting gravitational waves, where the exterior spacetime is described by
Robinson-Trautman geometries. The full nonlinear regime is examined by using the
Galerkin method that allows us to reduce the equations governing the dynamics to
a finite-dimensional dynamical system, after a proper truncation procedure.
Amongst the obtained results of the nonlinear evolution, one of the most
impressive is the fact that the distribution of the mass fraction extracted by
gravitational wave emission satisfies the distribution law of
nonextensive statistics and this result is independent of the initial
configurations considered.
\end{abstract}

\maketitle

Robinson-Trautman (RT) metrics are the simplest known
solutions of vacuum Einstein's equations which may be interpreted as
representing an isolated gravitational radiating system\cite{rt}. The
metric can be expressed as

\begin{eqnarray}
\label{eq1}
ds^2&=&\left(\lambda(u,\theta)+\frac{B(u)}{r}+2 r
\frac{\dot{K}}{K}\right) d u^2+2du dr -
r^{2}K^{2}(u,\theta)(d \theta^{2}+\sin^{2}\theta d \varphi^{2}),
\end{eqnarray}

\noindent where $r$ is an affine parameter defined along the shear-free null
geodesics determined by the vector field $\partial/\partial r$, dot means
derivative
with respect to $u$; $\lambda(u,\theta)=\frac{1}{K^2}-\frac{K_{\theta
\theta}}{K^3}+\frac{K_{\theta}^{2}}{K^4}-\frac{K_{\theta}}{K^3}\cot \theta$ is
the Gaussian
curvature of the surfaces $(u=const,r=const.)$. The geometry is nonstationary
and axially symmetric, admitting the obvious Killing vector
$\partial/\partial \varphi$. As a consequence of Einstein equations, we have the
following
evolution equation

\begin{equation}
\label{eq2}
-6 m_{0}\frac{\dot{K}}{K}+\frac{(\lambda_{\theta} \sin \theta)_{\theta}}{2 K^2
\sin \theta}=0.
\end{equation}

\noindent In the above the subscript $\theta$ denotes derivatives with
respect to $\theta$. Eq. (\ref{eq2}) is denoted the RT
equation, governs the dynamics of the gravitational field and will be the basis
of our analysis of the gravitational wave emission processes in RT spacetimes.
Formally speaking, it allows to evolve initial data $K(u,\theta)$ prescribed on
a given null surface $u=u_{0}$ (except in the case $m_{0}=0$). A particular and
important solution of the field equations is the Schwarzschild
metric obtained when $K(u,\theta)=K_{0}=const$, where
$\lambda(u,\theta)=K_{0}^{-2}$ and the corresponding Schwarzschild mass is $
M_{Schw}=m_{0}K_{0}^{3}$. This expression will be of particular importance in
our characterization of the mass function of the configuration. Concerning this
point, we note that in general the function $M(u,\theta)=B(u)K^{3}(u,\theta)$
is invariant under the coordinate transformation that reduces $B(u)$ to a
constant $-2m_{0}$; also this transformation induces that the type-D
${\cal{O}}(1/r^{3})$ curvature scalar associated with the mass aspect of the
spacetime has a $(u,\theta)$ dependence given exactly by $BK^3$.


Foster and Newman\cite{fn} exhibited the linearized solution of the field
equations. We generalize
their approach in a rather
distinct way in order to take into account the nonlinearities of the field
equations. For this proposal, we shall consider the Galerkin projection
method\cite{galerkin} in which we adopted the following
decomposition

\begin{equation}
K^2(u,x) = A_0^2\,{\rm e}^{Q(u,x)} = A_0^2\,{\exp}\left(\sum_{k=0}^{N}\,b_k(u)
P_k(x)\right), \label{eq3}
\end{equation}

\noindent where we have introduced a new variable $x = \cos \theta$ such that
$-1 \leq x \leq 1$; $N$ is the order of the truncation, $b_k(u)$ are the modal
coefficients, and we have adopted the Legendre polynomials $P_k(x)$ as the
basis functions of the projective abstract space. The internal product defined
in this abstract space is  $\left<P_{j}(x),P_{k}(x)\right> \equiv
\int_{-1}^{1}\,P_{j}(x) P_{k}(x)\,d x =
\frac{2\,\delta_{kj}}{2 k+1}$. The evolution equations for the modal
coefficients $b_k(u)$ are
derived in following way: we substitute (\ref{eq3}) into the equation
for $\lambda(u,x)$, both into (\ref{eq2}), and finally projecting the resulting
equation into each basis function $P_k(x)$. Then,

\begin{equation}
\dot{b}_{n}(u) =  \frac{(2 n+1)}{12 m_0
A_0^2}\,\left<{\rm
e}^{-Q(u,x)}\,[(1-x^2)\,\lambda^{\prime}]^{\prime},P_n(x)\right>.
\label{eq4}
\end{equation}

\noindent with $n=0,1,2..,N$. Therefore, the dynamics of RT spacetimes is
reduced to this system of ($N+1$) equations.

The initial conditions $b_k(u_0)$ are determined by the initial
data $K(u_0,x) \equiv k(x)$ since $b_j(u_0)=\frac{2 \left<\ln
k(x),P_j\right>}{\left<P_j,P_j\right>}$, $j=0,..,N$. We integrate numerically
the set of equations (\ref{eq4}) for distinct families $k(x)$ and the
following basic features of the nonlinear dynamics of RT spacetimes emerged. (i)
All modal coefficients tend to zero asymptotically; the exception is $b_0(u)$
that approaches to a constant value which we denote by $b_0(\infty)$. This
asymptotic configuration actually corresponds to the Schwarzschild solution with
mass $M_{\infty} = m_0 A_0^3\,\exp(3 b_0(\infty)/2)$.  Indeed, such a feature
feature is in agreement with previous analytical studies on RT
spacetimes\cite{chru}. (ii) For small values of the initial mass of the
distribution, $M_{init}$, more rapidly the Schwarzschild solution is settled
down; in addition a large fraction of $M_{init}$ is lost in the process due
gravitational wave extraction. On the opposite side, if $M_{init}$ is large,
then the evolution of RT spacetime towards the asymptotic Schwarzschild final
state is very slow. Correspondingly, the fraction of $M_{init}$ lost is very
small. (iii) The invariant characterization of the presence of gravitational
radiation can be done using the Peeling theorem. From this the complete pattern
gravitational wave emission is obtained\cite{deolso}. (iv) Finally, the last
important
feature is the relation between the fraction of mass extracted
$\Delta = \frac{|M_{init}-M_{\infty}|}{M_{init}}$ versus the initial mass
$M_{init}$. For
\textit{any} family $k(x)$ the mentioned relation satisfies the distribution
provided by the nonextensive statistics\cite{tsallis}.

\begin{figure}[ht]
\rotatebox{270}{\includegraphics*[scale=0.4]{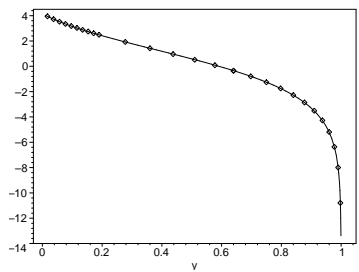}}
\caption{Log-linear plot of $\Delta$ versus $M_{init}$. The solid line corresponds
the nonextensive distribution
$\Delta = C_0\,(y_0-y)^{\alpha}\,[1+(q-1)\lambda_2\,(y_0-y)]^{1/1-q}$,
$y$ being associated to the initial mass, whereas the points were generated after integrating
the dynamical system (\ref{eq4}). The best fit is provided since $y_0=1 $, $C_0 \simeq 4.462$,
$\alpha \simeq 1.992$ and $q \simeq 1.725$. The initial data are ellipses with distinct
eccentricities.}
\end{figure}


The authors acknowledge the financial support of CNPq. H. P. de Oliveira is grateful to
ICTP where this work was initiated.

\end{document}